\documentclass[utf8]{FrontiersinHarvard} 

\usepackage{url,hyperref,lineno,microtype,subcaption}
\usepackage[onehalfspacing]{setspace}
\usepackage{comment}

\def\keyFont{\fontsize{8}{11}\helveticabold }
\def\firstAuthorLast{Li {et~al.}} 
\def\Authors{Yupei Li\,$^{1,+}$, Shuaijie Shao\,$^{2,+}$, Manuel Milling$^{3, 4}$, and Björn W.\ Schuller\,$^{1,3,4, 5*}$}
\def\Address{$^{1}$GLAM -- Group on Language, Audio, \& Music, Imperial College London, UK\\
$^{2}$University College London, UK\\
$^{3}$CHI -- Chair of Health Informatics, TUM University Hospital, Germany \\
$^{4}$ MCML -- Munich Center for Machine Learning, Germany \\
$^{5}$MDSI -- Munich Data Science Institute, Germany  \\
$^{+}$ Equal contribution
}

\def\corrAuthor{Yupei Li}

\def\corrEmail{yl7622@ic.ac.uk}

\begin{document}
\onecolumn
\firstpage{1}

\title[LLMs for Depression Recognition]{Large Language Models for Depression Recognition in Spoken Language Integrating Psychological Knowledge} 

\author[\firstAuthorLast ]{\Authors} 
\address{} 
\correspondance{} 

\extraAuth{}

\maketitle

\begin{abstract}

Depression is a growing concern gaining attention in both public discourse and AI research. While deep neural networks (DNNs) have been used for 
its 
recognition, they still lack real-world effectiveness.
Large language models (LLMs) show strong potential but require domain-specific fine-tuning and struggle with non-textual cues. Since depression is often expressed through vocal tone and behaviour rather than explicit text, relying on language alone is insufficient. Diagnostic accuracy also suffers without incorporating psychological expertise. To address these limitations, we present, to the best of our knowledge, the first application of LLMs to multimodal depression detection using the DAIC-WOZ dataset. We extract the audio features using the pre-trained model 
Wav2Vec, and mapped it to text-based LLMs for further processing. We also propose a novel strategy for incorporating psychological knowledge into LLMs to enhance diagnostic performance, specifically using a question and answer set to grant  authorised knowledge to LLMs. Our approach yields a notable improvement in both Mean Absolute Error (MAE) and Root Mean Square Error (RMSE) compared to a base score proposed by the related original paper. The codes are available in \href{https://github.com/myxp-lyp/Depression-detection.git}{Github}.
\tiny
 \keyFont{ \section{Keywords:} Large Language Models, Depression Recognition, Psychological Knowledge, Spoken Language, Speech} 
\end{abstract}

\section{Introduction}
As mental health gains increasing attention from the public, the diagnosis of emotional disorders -- particularly depression -- has become an essential area in both AI for healthcare and medical research. In previous decades, diagnosis has solely depended on the expertise of professional psychologists and clinicians. These circumstances introduce potential variability, as different clinicians may be influenced by their own intuition when assessing subjective questions, evident by \cite{gerber1989recognition}. To counteract this issue, diagnostic procedures are established with a maximum amount of standardisation, which stays limited despite major efforts. One opportunity to increase consistency and reproducibility of diagnosis is by leveraging automated assessments as a supporting tool, which is enabled by large-scale data and AI techniques 
\citep{zafar2024role}. A combination of AI,  
vast experiential data, and the professional knowledge of clinicians may together contribute to more accurate depression recognition, enhancing clinicians’ confidence and convenience, as well as increasing patient satisfaction, which is shown in the work of \cite{qassim2023mixed}. Although the use of AI alone to recognise depression has been proposed as a way to save time and reduce costs, patients have expressed skepticism and distrust, which \cite{robertson2023diverse} have revealed. Challenges include not only suboptimal detection rates but also the fact that current AI systems often lack integration with professional psychological 
knowledge, relying instead solely on data-driven experience.

Previous literature has made tremendous efforts to recognise depression within the domain of AI in healthcare. Two major research directions include the utilisation and adaptation of various deep neural networks (DNNs), as well as the fusion of multiple data modalities. A wide range of techniques for depression detection has been surveyed by \citet{squires2023deep}. Notably, \citet{valstar2013avec} introduced the depression detection challenge in 2013, laying the groundwork for subsequent research. Building on this, \citet{ringeval2019avec} proposed the use of Long Short-Term Memory (LSTM) models for the task, establishing a baseline for future approaches. Since then, various deep neural network (DNN) models have been developed, achieving promising results. For instance, \citet{niu2022dual} introduced Dual Attention and Element Recalibration Networks utilizing acoustic modalities for depression recognition.  
Despite their contributions, these models exhibit performance limitations largely due to the constrained learning capacity of their architectures. Recently, LLMs have demonstrated exceptional learning capabilities across many different fields (e.\,g., \citep{10111523}); however, their potential has not yet been effectively explored in the context of depression detection. Additionally, unimodal approaches are often insufficient for accurately assessing depression, as the evaluation process is inherently complex. Even clinicians frequently rely on indirect questioning during interviews to avoid triggering sensitive responses. Some prior studies have explored modality-specific models, such as LLM-based models focused on text~\cite{farruque2024depression}, autoencoder-based models centered on audio features~\cite{sardari2022audio}, and models targeting image and video modalities~ \cite{ashraf2020review}. These approaches, however, still lack definitive and comprehensive diagnostic information due to their unimodal nature. For instance, while audio may reveal weak pitch variations, the transcribed textual content might appear entirely normal, obscuring underlying emotional cues. Consequently, multimodal fusion is essential for a more holistic understanding of depressive symptoms. However, integrating multiple modalities into LLM-based models remains an open challenge, primarily because LLMs are mostly
trained on text-based tokens and are often not designed to natively process or fuse non-textual inputs.

Beyond technical challenges, it is also crucial for models to address concerns regarding patient trust. Current LLMs are not specifically trained on psychological or psychiatric knowledge comparable to the clinical experience of trained professionals. As a result, the responses generated by LLMs may lack authenticity and, more critically, may exhibit hallucinations -- a particularly serious issue in the context of depression recognition. Techniques such as knowledge injection, as discussed by  \cite{martino2023knowledge}, have been proposed to mitigate this limitation by integrating domain-specific information into LLMs. However, such methods have not yet been widely implemented in LLM-based models for depression recognition, leaving a considerable gap in ensuring both reliability and clinical validity.

Therefore, our paper proposes a novel approach to address the aforementioned research gaps:

\begin{itemize}
\item To the best of our knowledge, this is the first approach to directly apply LLMs for spoken language to the field of depression recognition. 
\item We introduce a pipeline that injects professional psychological knowledge into LLMs and demonstrates its effectiveness through empirical evaluation.
\item Our proposed pipeline considerably outperforms baseline models on the DAIC-WOZ dataset. 
\end{itemize}

The remainder of this paper is organised as follows. Section~\ref{sec:related} reviews related work in depression recognition, particularly focusing on DNNs and LLMs in a fusion of multiple streams context. Section~\ref{sec:dataset} describes the dataset used in our study, including its composition and its pre-processing. Section~\ref{sec:model} details our proposed multimodal LLM-based pipeline and the methodology for injecting psychological knowledge into the model. Section~\ref{sec:res} presents the experimental results, comparing our approach against established baselines. Finally, Section~\ref{sec:con} concludes the paper and outlines potential directions for future research.
\section{Materials and methods}

\subsection{Related work}
\label{sec:related}

\subsubsection{Text-based Depression Detection}
In recent years, Natural Language Processing (NLP) methods have increasingly utilised deep language models to identify depression symptoms in text. This approach is grounded in clinical practices, where mental health professionals often assess linguistic cues -- such as expressions of hopelessness, self-deprecation, or withdrawal -- to diagnose depression. In AVEC 2013, \citet{valstar2013avec} incorporated text modality for the first time, providing ASR transcripts alongside audio and video for multimodal depression detection and emotion recognition. Further, \cite{ogunleye2024sentiment} applied multiple hybrid models to two social media datasets, each involving binary classification tasks distinguishing between `depressed' and `not depressed' posts. Their combination of Sentence-BERT and an ensemble model achieved F1 scores of 69\% and 76\% on the respective datasets, demonstrating that incorporating lexicon-based sentiment indicators can enhance the performance of text-based models. This demonstrates that incorporating lexicon-based sentiment indicators can enhance the performance of text-based models. Similarly, \cite{s-etal-2022-scubemsec}, using a social media dataset from the Language Technology for Equality, Diversity, and Inclusion (LT-EDI) 2022 task published by the  Association for Computational Linguistics (ACL), trained several transformer models such as DistilBERT, ALBERT, and RoBERTa. Posts in the dataset were categorised as ``not depressed," ``moderately depressed," or ``severely depressed." Among the tested models, RoBERTa performed best, achieving an overall F1 score of 0.457 in the three-class problem, illustrating the effectiveness of transformer architectures for text-based depression classification tasks.
These studies are good examples of the current trend in research, in which researchers are increasingly focused on fine-tuning pretrained language models with labelled text. Previous studies also show that combining deep learning approaches with linguistic markers or psycholinguistic lexicons can substantially improve performance (e.\,g., \citep{10.3389/fpsyt.2023.1121583}). \cite{kathan2022journaling} proposed other format of text-based features such as behavioral activation for depression scale–short form (BADSSF), the center for epidemiologic studies depression scale (CESD), or the personality dynamics diary (PDD). Overall, the trend in text-based depression detection has shifted toward transformer-based models, often emphasising specific lexical indicators or sentence patterns related to emotional states. 

\subsubsection{Audio-based Depression Detection}
Depression symptoms can also manifest in vocal expression, prompting researchers to fine-tune pretrained speech models to capture acoustic features. For instance, individuals experiencing depression often exhibit paralinguistic characteristics such as reduced pitch variability, slower speech rate, and longer pauses. These features reflect the low arousal and negative emotional states 
often associated with depression. In AVEC 2013 \cite{valstar2013avec}, the first audio-based depression detection challenge, incorporated the audio modality as a key component for depression detection.
Moreover, \cite{huang2024depression} applied wav2vec 2.0 to the AVEC2017 dataset to extract audio features, achieving 96.5\% accuracy in binary depression classification. This highlights the capability of such models to learn high-quality representations without requiring complex processing. \cite{mallol2024multi} applied multi-triplet loss-based models for categorical depression recognition with four acoustic features. 
In addition, classical acoustic analysis remains effective. For example, \cite{berardi2023relative} extracted voice pathology features from recordings of picture descriptions and used them to train SVM classifiers. A third-degree polynomial SVM achieved over 92\% accuracy across all tasks. Their study identified articulatory precision, pause frequency, and speech variability as the most influential features. These represent two primary approaches in audio-based depression detection: (1) fine-tuning deep models like wav2vec and (2) extracting and classifying key acoustic features using traditional methods like SVM. Both approaches have proven effective for this task.

\subsubsection{Multimodal Depression Detection}
Multimodal models have been introduced into depression detection to combine textual, audio, and visual information, enabling richer feature representation than unimodal approaches. A prominent benchmark used in this field is the AVEC 2017 ``Real-life Depression and Affect Recognition'' challenge, which provides video, audio, and transcript data from interviews. \cite{Ringeval2017} introduced the according dataset with including PHQ-8 score regression. 
The baseline model for depression severity estimation contains text, audio, video, and combined audio and video models, serving as a reference for future work. Several studies have built upon this benchmark to improve it. 
\cite{sadeghi2024harnessing} extracted textual and facial-expression features using a LLM and a vision model, combining them to predict PHQ-8 scores. Their multimodal model slightly outperformed the text-only version in terms of mean squared error (MSE). Meanwhile, \cite{min2023detecting} collected annotated YouTube vlogs and conducted statistical analysis to highlight differences in depressive vs non-depressive videos. Their model learnt from both audio and video cues and achieved an F1 score of 77\% on the vlog dataset. Additionally, \cite{he2022depression} proposes a novel multimodal dataset collected from phone sensors, including phone calls, phone usage, and user activity. Similarly, \cite{kathan2022personalised} utilizes mobile sensors to collect multimodal data. These studies demonstrate that integrating multiple modalities -- whether text with facial expressions or audio with video -- can enhance model performance in depression recognition  tasks.

\subsubsection{LLM-based Depression Recognition}
The emergence of LLMs has brought major advancements for many application scenarios of AI including depression recognition, outperforming conventional non-large DNNs,
as were used before the rise of large models. \citet{schuller2024affective} show that LLMs have emotional emergence. Additionally, \citet{shin2024using} demonstrated the effectiveness of LLMs by prompting GPT-3.5 and GPT-4 with 428 diaries from 91 users to assess depression risk. With simple prompt engineering and minimal fine-tuning, the model achieved an accuracy of 90.2\% on binary depression classification, where a PHQ-9 score greater than 10 was considered indicative of depression. 
Notably, the fine-tuned GPT-3.5 outperformed its zero-shot untuned counterpart,
underscoring the potential of LLMs when adapted properly.
Expanding on this, \cite{liu2024emollms} introduced ``EmoLLMs,'' a series of LLMs fine-tuned for affective tasks. Trained on a multi-task emotional dataset, EmoLLMs surpassed GPT-4 on standard benchmarks, further showing how LLMs can be tailored to emotional understanding.
To assess how closely LLMs resemble human performance, \cite{zhang2023refashioningemotionrecognitionmodelling} compared models like ChatGPT, Claude, and Bing Chat on sentiment intensity tasks. 
Results showed that GPT-4 achieved scores comparable to humans, suggesting that LLMs can replicate or even exceed human-level emotional judgment. Together, these works indicate that LLMs represent a major leap forward in the field, with vast potential when appropriately fine-tuned.

\subsubsection{Knowledge-based Injection into LLMs}
Building on the strong baseline performance of LLMs, recent research has explored the integration of psychological knowledge to further improve mental health inference. \cite{li2025neuroplasticityartificialintelligence} surveyed the effectiveness of continuous learning, where knowledge-based injection as one potential approach. Specifically, \cite{abbasi2024psycholexunveilingpsychologicalmind} introduced ``PsychoLexLLaMA," an LLM designed for psychological assessment. Trained on the PsychoLex Q\&A dataset, this model learnt specialised psychological knowledge and outperformed other LLMs in reasoning tasks related to psychology, highlighting the benefits of domain-specific knowledge injection.
In another example, \cite{lan2024depression} proposed DORIS, a depression recognition system that incorporates clinical diagnostic knowledge. The model first used an LLM to identify social media posts containing DSM-related expressions, then generates emotional summaries and estimates emotional intensities. These outputs are fed into a conventional classifier, resulting in improved performance for binary depression classification (depressed vs control) compared to standard models.
Similarly, \cite{tank2024depression} demonstrated that encoding questionnaire knowledge into prompts enhances LLM effectiveness. Their system for PHQ-8 scoring uses structured prompts based on depression symptoms and a two-shot classification setup. They found that embedding relevant questionnaire knowledge increased prediction accuracy.
Together, these studies reveal that enriching LLMs with psychological or diagnostic knowledge can further elevate their capability in depression recognition tasks.

\subsection{Dataset and Preprocessing}
\label{sec:dataset}
\subsubsection{Experimental Dataset}
The dataset
used in this study is DAIC-WOZ \citep{gratch-etal-2014-distress}. The interviews feature a patient interacting with a virtual human interviewer named Ellie (as indicated in the transcript files). The dataset includes audio recordings of complete conversations and their corresponding transcripts, which specify the speaker, the spoken content, and the start and end times of each utterance. Although the dataset also contains facial feature data, this was not utilised in the current research.

DAIC-WOZ consists of 189 samples, also referred to as 189 participants, pre-divided into training, validation, and test sets, containing 107, 35, and 47 samples respectively. Additionally, a metadata table is provided, which includes patient IDs, binary depression labels, PHQ-8 scores, and dimensional depression scores (the latter were not used in this study).

\subsubsection{Data Preprocessing}
To enable the LLM 
to better interpret and extract features from the audio data, the recordings were segmented to accommodate the limited context window of LLMs. Using the transcript file timestamps and speaker annotations, each sentence spoken by a  participant was isolated. Every five consecutive utterances were then merged into a single audio file, ensuring that no audio segments from different participants were combined. 
Given there is a ``speaker" value in the dataset, corresponding transcript data was filtered to include only the speech of the participant. The sentences were merged in the same manner as the audio: every five sentences were combined into one element in the new set, subsequently  merging the elements into a CSV file. Some useless values were eliminated in this file, keeping only the merged text content, the associated participant ID. Note that the dataset only provides a single sentence-independent PHQ-8 score per participant, which we therefore take as the target for all sentences from the same speaker.

After preprocessing, a total of 6,556 audio files were generated, 
each with a corresponding transcript entry, as both were segmented using the same criteria. Each file obtains its label in form of the participant’s PHQ-8 score. During the final prediction stage, the model generates a score for each individual audio segment. These segment-level predictions are then averaged to produce the overall PHQ-8 score for the corresponding participant.

\subsection{Model Pipeline}
\label{sec:model}
\begin{figure}[ht!]
\begin{center}
\includegraphics[width=\linewidth]{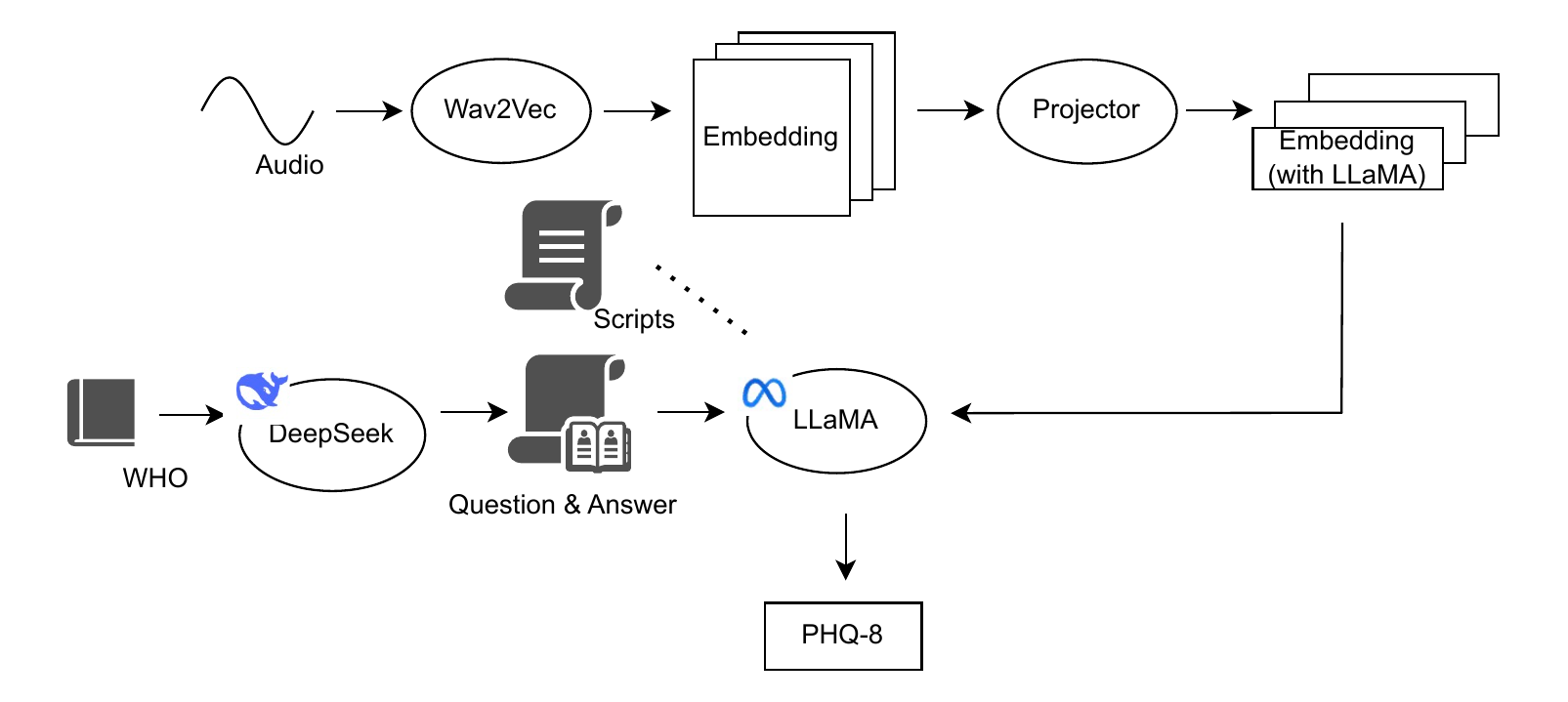}
\end{center}
\caption{Two-Stage Pipeline of the Large Models considered. Our proposed framework consists of two key stages. In the first stage, we leverage the DeepSeek model to extract question–answer pairs from authoritative psychology texts, such as disease definitions and clinical descriptions from the World Health Organization (WHO). These extracted pairs, along with transcript data, are used to pretrain the LLM through a process of knowledge injection, enhancing its domain-specific understanding. In the second stage, we process audio inputs using a feature-extraction DNN, Wav2Vec to obtain their original embeddings. These embeddings are then projected via a feedforward network to align them with the LLM's hidden space. Once the audio and text modalities are integrated, the LLM, LLaMA is used to predict the depression level.}\label{fig:pipeline}
\end{figure}

We have organised our model pipeline as illustrated in Figure~\ref{fig:pipeline}. We selected the acoustic information as the primary input. 
The acoustic signal is directly  available during clinical interviews, whereas text-based methods require an additional transcription step such as by automatic speech recognition. Nevertheless, the spoken text remains valuable for providing clear and explicit content regarding the subject's verbal expressions. Therefore, we incorporate both types of information to complement each other and enhance overall performance.

We split our LlaMA 
fine-tuning with two main 
phases, namely knowledge injection and PHQ-8 score prediction. 
\subsubsection{Psychology Knowledge Injection}
To address the lack of domain-specific knowledge in LLMs, we design a learning process that mirrors human cognitive development. Specifically, we aim for the LLM to read and comprehend psychological knowledge in a manner similar to how humans study and internalise the information. Inspired by the work of \cite{abbasi2024psycholexunveilingpsychologicalmind}, we extract question–answer pairs from psychological sources and prompt the LLM to generate responses accordingly. In contrast to \cite{abbasi2024psycholexunveilingpsychologicalmind}, our approach introduces more structured and comprehensive question types to facilitate deeper understanding. Drawing from principles of human learning outlined in \cite{novak1984learning}, effective learning requires understanding what the knowledge is, why it matters, how it is applied, and how it connects with prior knowledge to foster critical thinking and practical use. In line with this, we design six distinct types of questions centred on depressive disorders: (1) the definition of the disorder, (2) the rationale for diagnosing it, (3) common symptoms or manifestations, (4) extended or related knowledge, and (5) critical thinking questions. This structured framework ensures that the LLM not only memorises facts but also develops a more nuanced and applicable models of depression.

We selected the World Health Organization's official medical classification website\footnote{\url{https://icd.who.int/browse/2024-01/mms/en}} as our primary knowledge source due to its authoritative content, which helps reduce the risk of incorporating low-quality or misleading information that could contribute to hallucinations in LLMs. To focus specifically on depression-related knowledge, we extracted a subset of entries by filtering disorder titles using relevant keywords (e.\,g., anxiety, 
depress*, mood, stress, chronic, isolation), resulting in a total of 123 samples. We then employed the DeepSeek-V3 model from \cite{liu2024deepseek} to generate structured question–answer pairs based on the provided content. Specifically, we used the following prompt to guide the generation process: 
\begin{quote}
\itshape
Construct Q\&A sets based on one [paragraph] I give you. 

(1) The 10 Q\&A sets should be about the key definitions mentioned in the paragraph. The Q\&A set should contain no extra knowledge.
(2) The 10 questions in these sets should be `why' questions. The Q\&A set should contain no extra knowledge.
(3) The 10 Q\&A are about the phenomena that may occur on people with such disorder. The Q\&A set should contain no extra knowledge.
(4) The 5 Q\&A sets should be completely based on extended knowledge which is not mentioned in the [paragraph], but should also be considered important about such disorder.
(5) The five Q\&A sets should show critical thinking. The Q\&A set should contain no extra knowledge.

The entire conversation should contain English only. The message you reply must follow the exact format in the [example], do not add any extra \texttt{"} or other marks at the beginning or the end of your question or answer.

[example]:\\
question: This is the first question you construct.\\
answer: This is the first answer you construct.
\end{quote}
This process resulted in a total of 4,920 question–answer pairs. We then provided the questions as prompts to our LLM, LLaMA \cite{touvron2023llamaopenefficientfoundation}\footnote{\url{https://huggingface.co/meta-llama/Llama-2-7b-hf}}, and employed supervised learning to train the model to generate corresponding answers. Through this process, we aim to effectively inject psychological knowledge into the model, enabling it to better understand and respond to depression-related content.
\subsubsection{Multi-stream Training}
To train the model with multi-stream features, we trained the LLM on text features first using prompts:
\begin{quote}
\itshape
Transcripts:[Transcript], PHQ Score:
\end{quote}
This enables LLM to learn text features first.

Next, to train with audio features and align them with the text-based embedding space of the LLM, we aim to project the audio representations into a shared latent space of the LLM. Inspired by SALMONN, a speech-based LLM proposed by \cite{tang2024salmonngenerichearingabilities}, we adopt a similar pipeline but with a modification: instead of using Whisper \cite{radford2022robustspeechrecognitionlargescale} to extract audio features, which is primarily designed for speech recognition tasks, we utilise Wav2Vec 2.0 \cite{baevski2020wav2vec20frameworkselfsupervised}. We selected Wav2Vec 2.0 due to its stronger capacity for capturing a broader range of audio features, which are more relevant for understanding emotional cues and prosodic elements associated with depression. These have been formulated below.
\begin{align}
    r &= \text{Wav2Vec2}(raw_{audio}) \\
    Emb_{audio} &= feedforward(r) \\
    \text{PHQ-8} &= \text{Linear}(\text{LLaMA}(Emb_{audio})_{-1}),
\end{align}
where $r$ is in shape of $R^{s*d_a}$, with $s$ being extracted audio length by Wav2Vec2 and $d_a$ being the hidden dimension, and $Emb_{audio}$ is in shape of $R^{s*d_t}$, with $d_t$ being the hidden dimension of LLaMA. In this manner, the audio features are effectively mapped and projected into the embedding space of the LLaMA model, enabling seamless integration with its text-based representations. This allows LLaMA to learn meaningful representations for the audio modality as well. Finally, we utilise the last hidden layer embedding and use a linear layer to predict the PHQ-8 score.

\subsubsection{Experiments}
To ensure reproducibility, we fixed the random seed to 42 using \emph{random} library in Python for \emph{Numpy} and \emph{torch} usage.
Due to limited GPU resources, we employed Low-Rank Adaptation (LoRA) \cite{hu2021loralowrankadaptationlarge} with Distributed Data Parallel (DDP) for efficient fine-tuning of the LLM. The detailed hyperparameter settings used during training are summarised in Table~\ref{tab:hyper}.

\begin{table}[h!]
\centering
\begin{tabular}{c|c}
\hline
\textbf{Parameter} & \textbf{Value} \\
\hline
$r$ & 8 \\
\hline
\texttt{lora\_alpha} & 16 \\
\hline
\texttt{lora\_dropout} & 0.1 \\
\hline
\texttt{target\_modules} & \texttt{``q\_proj", ``v\_proj"]} \\
\hline
\texttt{GPU} & Tesla V100-PCIE-32GB \\
\hline
\end{tabular}
\caption{LLaMA hyper-parameters}
\label{tab:hyper}
\end{table}

\section{Results}
\label{sec:res}
We conducted a series of experiments, and the results are presented in Table~\ref{tab:results} including ablation studies to measure the impact of certain components of our method. 
To evaluate model performance, we used Mean Absolute Error (MAE) and Root Mean Square Error (RMSE) as the primary metrics. 
Additionally, the final prediction scores are calculated by each participant across multiple clips derived from the same individual.
\begin{table}[h!]
\centering
\begin{tabular}{l|c|c}
\hline
\textbf{Model} & \textbf{MAE}& \textbf{RMSE} \\
\hline
AVEC2016 (audio) \cite{valstar2016avec2016depression}& 5.72 & 7.78 \\
LSTM \cite{afzal2023automated}& 5.7 & 6.59 \\
Random Forest \cite{afzal2023automated}& 5.71 & 6.79 \\

\hline
Ours(audio) & 5.373 & 6.733 \\
Ours(text) & 6.342 & 8.891 \\
Ours(audio + text) & 5.356 & 6.713\\
Ours(text + Knowledge Injection) & \textbf{5.354} & \textbf{6.429} \\
Ours(audio + Knowledge Injection) & 5.356 & 6.713\\
Ours(audio +text+ Knowledge Injection) & 5.356 & 6.713\\

\hline
\end{tabular}
\caption{Evaluation on DAIC-WOZ Test set for different models}
\label{tab:results}
\end{table}

Our pure audio-based method shows a clear improvement over the baseline, 
which is only available for audio in the AVEC 2016 challenge (rather than also for text or audio + text) \cite{valstar2016avec2016depression}.
This demonstrates that the audio features in the model contribute to a better performance, suggesting that audio signals provide valuable information for the task, outperforming the baseline in both the MAE and RMSE metrics.

Additionally, knowledge injection has a noticeable impact on LLM depression prediction. It lowers both MAE and RMSE marginally compared to the plain audio and text model. The MAE drops from 5.373 to 5.356 for audio, from 6.342 to 5.354 for text, and RMSE drops from 6.733 to 6.713 for audio, from 8.891 to 6.429, for text. The performance improves more for text than for audio. This suggests that the text-based knowledge injection has a stronger effect, likely because the knowledge is provided in a textual form, which aligns better with the text modality.

When combining both audio and text with knowledge injection, the performance is close to text with knowledge injection, and pure text with audio.  While this suggests the potential for combining the information streams, it also highlights that the combination may not always result in substantial performance gains, possibly due to the fact that the learning ability of LLMs is biased towards audio. Additionally, it may due to the LLMs learning ability is limited by small number of trainable parameters from LoRA. However, the combination still performs well and indicates that both modalities can be useful when leveraged appropriately.

Taking another perspective, we investigate whether the knowledge injection works as expected. The procedure involves instructing PsycholexLLAMA to generate fifty questions. The prompt used is relatively straightforward.
\begin{quote}
    \itshape
    Regard yourself as a teacher teaching about depression detection, and you are trying to test how well your students understand this knowledge. Generate {num\_questions} specific questions about psychological assessment of depression to evaluate whether the students know how to assess depression. Number them as 1., 2., 3., etc.
\end{quote}
The generated questions were then submitted to two different models for response, the one trained with the Question \& Answer pairs and the one without such post-training. 
Both the questions and the corresponding answers were subsequently passed back to PsycholexLLAMA, which was tasked with scoring each response on a scale from 0 to 10. Specifically, we adopted this prompt:
\begin{quote}
    \itshape
    Rate the following response to a question about depression on a scale from 0-10:

                QUESTION: {question}

                RESPONSE: {response}

                Evaluate based on:
                - Accuracy of information
                - Sensitivity to mental health issues
                - Helpfulness of the response
                - Appropriate tone

                Rate from 0 (completely inappropriate) to 10 (excellent in all aspects).
                Your rating (just the number):
\end{quote}
For each question, the higher of the two scores was recorded, which follows the principle of the pass@2 strategy to measure performance of LLMs proposed by \cite{chen2021evaluating}. Finally, the average of these highest scores was calculated to yield the overall performance score. The original model achieved a score of 7.32, whereas our trained model obtained a score of 8.20, indicating that our proposed strategy is effective and that the model has successfully learnt from the psychology knowledge base.

While our strategy demonstrates promising results, it also highlights two key directions for future work. First, although LoRA offers a lightweight and efficient fine-tuning method, its capacity to induce substantial changes in model behaviour is limited compared to full-scale training of LLMs. However, fully training an LLM demands significant computational resources, which may not always be feasible. Second, as previously discussed, there is a notable scarcity of psychologically-informed audio data. As a result, the model -- despite being equipped with theoretical knowledge derived from text -- struggles to effectively transfer this understanding to audio-based applications. Generating such audio content artificially using AI can often lack authenticity. A potential solution would be the development of clinically annotated audio datasets that pair spoken examples with corresponding textual explanations -- such as diagnostic insights based on criteria from the WHO. Such resources could serve as a valuable bridge between theory and practical application in audio-based psychological assessments.
\section{Conclusion}
\label{sec:con}
In conclusion, our work presented a novel approach to advancing depression detection by leveraging the capabilities of LLMs. By incorporating professional psychological knowledge into the considered LLM through a carefully designed pipeline, we enhanced the model's ability to interpret and evaluate depressive symptoms more effectively. Our empirical results, validated on the DAIC-WOZ 
dataset, demonstrate that the proposed method substantially outperforms established baselines (around 0.35 on MAE and 1.36 on RMSE), underscoring the potential of LLMs as a powerful tool in mental health assessment. These findings pave the way for future research exploring the integration of domain expertise into multimodal AI systems for clinical applications. In the future, the work may continue to fully train LLMs and obtain annotations or suggestions from specialists. Additionally, future work should explore training full-scale LLMs and incorporating richer sound descriptions as a form of knowledge injection, in order to better bridge the gap between acoustic features and the text-based knowledge learned by LLMs.
\section*{Conflict of Interest Statement}

The authors declare that the research was conducted in the absence of any commercial or financial relationships that could be construed as a potential conflict of interest.

\section*{Author Contributions}

Yupei Li -- all CRediT roles

Shuaijie Shao -- Data curation, Formal analysis, Investigation, Software, Writing -- original draft

Manuel Milling -- Supervision (discussing ideas), Validation, Funding acquisition, Writing -- review \& editing

Björn W.\ Schuller -- Supervision (main supervisor), Validation, Funding acquisition, Writing -- review \& editing


\section*{Funding}
This research was partially supported and funded by the Munich Center for Machine Learning and the Munich Data Science Institute.


\section*{Acknowledgments}
We acknowledge Hanqian Li from Shandong University for providing initial draft experiment codes, Adria Mallol Ragolta from Technical University Munich to discuss ideas with us.

\section*{Data Availability Statement}
The datasets DAIC-WOZ for this study can be found  \href{https://dcapswoz.ict.usc.edu/}{online}

\bibliographystyle{Frontiers-Harvard} 
\bibliography{test}



\end{document}